\documentclass[draft,showpacs,eqsecnum,nofootinbib,aps]{revtex4}
\renewcommand{\theequation}{\arabic{equation}}
\def\beq{\begin{equation}}
\def\eeq{\end{equation}}
\def\bea{\begin{eqnarray}}
\def\eea{\end{eqnarray}}
\def\nn{\nonumber}

\begin{document}
\title{SO(3,2)/Sp(2) symmetries in BTZ black holes}

\author{Soon-Tae Hong}
\email{soonhong@ewha.ac.kr}
\affiliation{Department of Science
Education, Ewha Womans University, Seoul 120-750 Korea}
\date{October 8, 2003}

\vskip3.0cm
\begin{abstract}
We study global flat embeddings, three accelerations and Hawking temperatures
of the BTZ black holes in the framework of two-time physics scheme associated
with Sp(2) local symmetry, to construct their corresponding SO(3,2) global symmetry
invariant Lagrangians both inside and outside event horizons.
Moreover, the Sp(2) local symmetry is discussed in terms of the
metric time-independence.
\end{abstract}
 \maketitle

\section{Introduction}
\setcounter{equation}{0}
\renewcommand{\theequation}{\arabic{section}.\arabic{equation}}
There have been tremendous progresses in lower dimensional
black holes associated with the string theory since an exact conformal field theory
describing a black hole in two-dimensional space-time was proposed~\cite{witten91}.
Especially, the (2+1) dimensional
Banados-Teitelboim-Zanelli (BTZ) black holes~\cite{btz1,btz2,cal}
have enjoyed lots of successes in relativity and string communities, since
thermodynamics of higher dimensional black holes can be interpreted
in terms of the BTZ back hole solutions.  In fact, the dual solutions
of the BTZ black holes are related to the solutions in the string theory,
so-called (2+1) black strings~\cite{horowitz93,callan85}.

On the other hand, the higher dimensional global flat embeddings of the
black hole solutions are subjects of great interest both to mathematicians
and to physicists.  In differential geometry, it has been well-known that
four dimensional Schwarzschild metric is not embedded in
$R^{5}$~\cite{spivak75}.  Recently, (5+1) dimensional global embedding Minkowski
space structure for the Schwarzschild black hole has been obtained~\cite{deser97}
to investigate a thermal Hawking effect on a curved manifold~\cite{hawk75} associated with an
Unruh effect~\cite{unr} in these higher dimensional space time.  It has been also shown
that the uncharged and charged BTZ black holes
are embedded in (2+2)~\cite{deser97} and (3+2) dimensions,\footnote{This (3+2) minimal
embedding is constructed in Appendix.  In the previous
work~\cite{kps00}, the charged BTZ black hole has (3+3) nonminimal embedding
which is, however, reduced to the uncharged BTZ embedding in the vanishing charge limit.}
while the uncharged and charged black strings are embedded
in (3+1) and (3+2) dimensions~\cite{hong01prd}, respectively.  Note that
one can have two time coordinates in these embedding solutions to suggest
so-called two-time physics~\cite{bars98}.  Historically, the two-time physics
was formulated long ago when the (3+1) Maxwell theory on a conformally invariant
(4+2) manifold was constructed~\cite{dirac36}.  Recently, the two-time physics scheme
has been applied to  M theory~\cite{bars99} and noncommutative gauge theories~\cite{bars01}.

In this paper we will investigate symmetries involved
in the BTZ black hole embeddings such as SO(3,2) global and Sp(2) local symmetries.  In section 2,
we will study complete embedding solutions, three accelerations and Hawking temperatures ``inside
and outside" the event horizons in the framework of the two-time physics and then construct
the SO(3,2) global symmetry invariant Lagrangians
associated with these embedding solutions in section 3.  The Sp(2) local symmetry will also be
discussed in terms of the metric time-independence.  In Appendix, we will revisit the charged BTZ
black hole to construct its minimal embedding solution.

\section{Complete flat embedding geometries}
\setcounter{equation}{0}
\renewcommand{\theequation}{\arabic{section}.\arabic{equation}}

We first briefly recapitulate the global flat embedding solution given
in~\cite{deser97}, for the (2+1) rotating BTZ black hole~\cite{btz1,btz2} which is described
by 3-metric
\beq
ds^2=-N^2dt^2+N^{-2}dr^{2}+r^{2}(d{\phi}+N^{\phi}dt)^{2},
\label{3metricr}
\eeq
where the lapse and shift functions are
\beq
N^{2}=-M+ \frac{r^2}{l^2} + \frac{J^2}{4r^2},~~ N^{\phi}=-\frac{J}{2r^2},
\eeq
respectively. Note that for the nonextremal case there exist two horizons
$r_{\pm}(J)$ satisfying
the following equations,
\beq 0=-M+ \frac{r_{\pm}^2}{l^2}+\frac{J^2}{4r_{\pm}^2},
\eeq
in terms of which we can rewrite the lapse and shift functions as follows
\beq
N^2= \frac{(r^2 - r_{+}^2)(r^2 -r_{-}^2)}{r^2 l^2},~~
N^{\phi}=-\frac{r_{+}r_{-}}{r^2 l}.
\label{4eqns}
\eeq
Here one notes that this BTZ space originates from
Anti-de Sitter one via the geodesic identification $\phi=\phi+2\pi$.
The (2+2) minimal BTZ global flat embedding $ds^{2}=-(dz^{0})^2+(dz^{1})^2
+(dz^{2})^2-(dz^{3})^{2}$ is then given by the coordinate transformations
for $r\geq r_{+}$ as follows
\bea
z^{0}&=&l\left(\frac{r^{2}-r_{+}^{2}}{r_{+}^{2}-r_{-}^{2}}\right)^{1/2}
        \sinh
        \left(\frac{r_{+}}{l^2}t-\frac{r_{-}}{l}\phi\right), \nonumber \\
z^{1}&=&l\left(\frac{r^{2}-r_{+}^{2}}{r_{+}^{2}-r_{-}^{2}}\right)^{1/2}
        \cosh
        \left(\frac{r_{+}}{l^2}t-\frac{r_{-}}{l}\phi\right), \nonumber \\
z^{2}&=&l\left(\frac{r^{2}-r_{-}^{2}}{r_{+}^{2}-r_{-}^{2}}\right)^{1/2}
       \sinh\left(\frac{r_{-}}{l^2}t-\frac{r_{+}}{l}\phi\right), \nonumber \\
z^{3}&=&l\left(\frac{r^{2}-r_{-}^{2}}{r_{+}^{2}-r_{-}^{2}}\right)^{1/2}
       \cosh\left(\frac{r_{-}}{l^2}t-\frac{r_{+}}{l}\phi\right).
\label{btz2}
\eea
In the following we will construct complete embedding solutions, three accelerations and
Hawking temperatures ``inside and outside" the event horizons.

\subsection{Case I: $r\geq r_{+}$}

In the two-time physics scheme~\cite{bars98}, we consider global flat embedding
structure for the (2+1) rotating BTZ black hole whose metric is now given by
\bea
ds^{2}&=&\eta_{MN}dX^{M}dX^{N}
\label{2tmetric}\\
X^{M}&=&l(c_{-}\cosh R,s_{-}\cosh R,s_{+}\sinh R, c_{+}\sinh R, 1)
\label{xm1}
\eea
where $\eta_{MN}={\rm diag}~(-1,+1,-1,+1,+1)$ with $M=(0^{\prime},1^{\prime},0,1,2)$ and
\beq
c_{\pm}=\cosh
\frac{x^{\pm}}{l},~~~s_{\pm}=\sinh \frac{x^{\pm}}{l}.
\label{cpm}
\eeq
With the canonical momenta $p_{\mu}$ conjugate to $x_{\mu}$ given as
\beq
x^{\mu}=(\frac{x^{+}}{l},R,\frac{x^{-}}{l}),~~~
p_{\mu}=(p_{+},p_{R},p_{-}),
\label{xpmunu}
\eeq
we can construct
\bea
P^{M}&=&\frac{1}{l}(c_{-}p_{R}\sinh R +s_{-}p_{-}{\rm sech}R,s_{-}p_{R}\sinh R+c_{-}p_{-}{\rm sech} R,
\nonumber\\
& &s_{+}p_{R}\cosh R-c_{+}p_{+}{\rm cosech} R,c_{+}p_{R}\cosh
R-s_{+}p_{+}{\rm cosech} R,0).\nn\\
\label{pm1}
\eea
to satisfy the Sp(2) local symmetry associated with the two-time physics,\footnote{Here one
can easily show that $X_{M}X^{M}=X_{M}P^{M}=0$.  The last symmetry condition
$P_{M}P^{M}=0$ will be discussed later.}
\beq
X_{M}X^{M}=0,~~~X_{M}P^{M}=0,~~~P_{M}P^{M}=0.
\label{xmxmpm}
\eeq

Now we differentiate $X^{M}$ in (\ref{xm1}) to yield
\beq
ds^{2}=dX_{M}dX^{M}=l^{2}dR^{2}-\sinh^{2}R~(dx^{+})^{2}+\cosh^{2}R~(dx^{-})^{2},
\eeq
which,  using the identification for $x^{\pm}$
\beq
x^{\pm}=\frac{r_{\pm}}{l}t-r_{\mp}\phi,
\label{xpman}
\eeq
can be rewritten as for $r\geq r_{+}$
\bea
ds^{2}&=&l^{2}dR^{2}+l^{-2}(r_{-}^{2}\cosh^{2}R-r_{+}^{2}\sinh^{2}R)dt^{2}+2r^{2}N^{\phi}dtd\phi
\nonumber\\
& &+(r_{+}^{2}\cosh^{2}R-r_{-}^{2}\sinh^{2}R)d\phi^{2}.
\eea
Exploiting the ansatz \footnote{In the literature~\cite{von99}, there appears a brief
sketch on the BTZ embedding outside the horizon in the two-time physics scheme,
without explicit construction of $\cosh R$ and $\sinh R$ in (\ref{coshout})
for instance.} for $\cosh R$ and $\sinh R$
\beq
\cosh R=\left(\frac{r^{2}-r_{-}^{2}}{r_{+}^{2}-r_{-}^{2}}\right)^{1/2},~~~
\sinh R=\left(\frac{r^{2}-r_{+}^{2}}{r_{+}^{2}-r_{-}^{2}}\right)^{1/2}
\label{coshout}
\eeq
we can reproduce the BTZ metric (\ref{3metricr}) and the (3+2) global flat embedding for $r\geq r_{+}$
\bea
X^{0^{\prime}}&=&l\left(\frac{r^{2}-r_{-}^{2}}{r_{+}^{2}-r_{-}^{2}}\right)^{1/2}
       \cosh\left(\frac{r_{-}}{l^{2}}t-\frac{r_{+}}{l}\phi\right),\nonumber\\
X^{1^{\prime}}&=&l\left(\frac{r^{2}-r_{-}^{2}}{r_{+}^{2}-r_{-}^{2}}\right)^{1/2}
       \sinh\left(\frac{r_{-}}{l^{2}}t-\frac{r_{+}}{l}\phi\right), \nonumber \\
X^{0}&=&l\left(\frac{r^{2}-r_{+}^{2}}{r_{+}^{2}-r_{-}^{2}}\right)^{1/2}\sinh
        \left(\frac{r_{+}}{l^{2}}t-\frac{r_{-}}{l}\phi\right), \nonumber \\
X^{1}&=&l\left(\frac{r^{2}-r_{+}^{2}}{r_{+}^{2}-r_{-}^{2}}\right)^{1/2}\cosh
        \left(\frac{r_{+}}{l^{2}}t-\frac{r_{-}}{l}\phi\right), \nonumber \\
X^{2}&=&l,
\label{btz22}
\eea
to yield the identification between the standard global flat embedding (\ref{btz2}) and
that of two-time physics: $(X^{0^{\prime}},X^{1^{\prime}},X^{0},X^{1})=(z^{3},z^{2},z^{0},z^{1})$.
Here one notes that $X^{2}=l$ does not contribute the minimal global flat embedding
since it is constant, and this coordinate $X^{2}$ serves to fulfill the Sp(2) symmetry (\ref{xmxmpm}).

Next, introducing the Killing vector $\xi=\partial_{t}-N^{\phi}\partial_{\phi}$ we evaluate the three
acceleration \beq
a_{3}=\frac{r^{4}-r_{+}^{2}r_{-}^{2}}{r^{2}l(r^{2}-r_{+}^{2})^{1/2}
(r^{2}-r_{-}^{2})^{1/2}}, \eeq
and the Hawking temperature~\cite{hawk75}
\beq
T_{H}=\frac{a_{4}}{2\pi}=\frac{r(r_{+}^{2}-r_{-}^{2})}{2\pi
r_{+}l(r^{2}-r_{+}^{2})^{1/2} (r^{2}-r_{-}^{2})^{1/2}},
\label{htemp1}
\eeq
which are consistent with the fact that the $a_{4}$ is also attainable
from the relation~\cite{hawk75}
\beq
a_{4}=\frac{k}{g_{00}^{1/2}}.
\eeq

\subsection{Case II: $r_{-}\le r\le r_{+}$}

Next, we consider the global flat embedding of the BTZ black hole in the
range of $r_{-}\leq r\leq r_{+}$ by exploiting a little bit
different choice for $X^{M}$
\beq
X^{M}=l(c_{-}\cos R,s_{-}\cos R,c_{+}\sin R, s_{+}\sin R, 1)
\label{xm2}
\eeq
which satisfies the Sp(2) local symmetry (\ref{xmxmpm}).
As in the previous section, differentiating $X^{M}$ in (\ref{xm2}) yields for
$r_{-}\leq r\leq r_{+}$
\bea
ds^{2}&=&-l^{2}dR^{2}+l^{-2}(r_{+}^{2}\sin^{2}R+r_{-}^{2}\cos^{2}R)dt^{2}
+2r^{2}N^{\phi}dtd\phi
\nonumber\\
& &+(r_{+}^{2}\cos^{2}R+r_{-}^{2}\sin^{2}R)d\phi^{2}. \eea
Exploiting the ansatz for $\cos R$ and $\sin R$
\beq
\cos R=\left(\frac{r^{2}-r_{-}^{2}}{r_{+}^{2}-r_{-}^{2}}\right)^{1/2},~~~
\sin
R=\left(\frac{r_{+}^{2}-r^{2}}{r_{+}^{2}-r_{-}^{2}}\right)^{1/2}
\label{cosbetw}
\eeq
we can obtain the (3+2) global flat embedding for $r_{-}\leq r\leq r_{+}$
\bea
X^{0^{\prime}}&=&l\left(\frac{r^{2}-r_{-}^{2}}{r_{+}^{2}-r_{-}^{2}}\right)^{1/2}
       \cosh\left(\frac{r_{-}}{l^{2}}t-\frac{r_{+}}{l}\phi\right),\nonumber\\
X^{1^{\prime}}&=&l\left(\frac{r^{2}-r_{-}^{2}}{r_{+}^{2}-r_{-}^{2}}\right)^{1/2}
       \sinh\left(\frac{r_{-}}{l^{2}}t-\frac{r_{+}}{l}\phi\right), \nonumber \\
X^{0}&=&l\left(\frac{r_{+}^{2}-r^{2}}{r_{+}^{2}-r_{-}^{2}}\right)^{1/2}
        \cosh
        \left(\frac{r_{+}}{l^{2}}t-\frac{r_{-}}{l}\phi\right), \nonumber \\
X^{1}&=&l\left(\frac{r_{+}^{2}-r^{2}}{r_{+}^{2}-r_{-}^{2}}\right)^{1/2}
        \sinh
        \left(\frac{r_{+}}{l^{2}}t-\frac{r_{-}}{l}\phi\right), \nonumber \\
X^{2}&=&l. \label{btz222} \eea

Next, similar to the $r\geq r_{+}$ case, with the Killing vector
$\xi=\partial_{t}-N^{\phi}\partial_{\phi}$ we evaluate the three
acceleration \beq
a_{3}=\frac{r^{4}-r_{+}^{2}r_{-}^{2}}{r^{2}l(r_{+}^{2}-r^{2})^{1/2}
(r^{2}-r_{-}^{2})^{1/2}}, \eeq and the Hawking temperature
\beq
T_{H}=\frac{a_{4}}{2\pi}=\frac{r(r_{+}^{2}-r_{-}^{2})}{2\pi
r_{+}l(r_{+}^{2}-r^{2})^{1/2} (r^{2}-r_{-}^{2})^{1/2}}.
\label{temp2}
\eeq

\subsection{Case III: $r\leq r_{-}$}

Similarly, for the case of the range inside the inner horizon,
$r\leq r_{-}$, we introduce \beq X^{M}=l(s_{-}\sinh R,c_{-}\sinh
R,c_{+}\cosh R, s_{+}\cosh R, 1)
\label{xm3}
\eeq
to yield
\bea
ds^{2}&=&l^{2}dR^{2}+l^{-2}(r_{+}^{2}\cosh^{2}R-r_{-}^{2}\sinh^{2}R)dt^{2}+2r^{2}N^{\phi}dtd\phi
\nonumber\\
& &+(r_{-}^{2}\cosh^{2}R-r_{+}^{2}\sinh^{2}R)d\phi^{2}.
\eea
With the ansatz for $\cosh R$ and $\sinh R$
\beq
\cosh R=\left(\frac{r_{+}^{2}-r^{2}}{r_{+}^{2}-r_{-}^{2}}\right)^{1/2},~~~
\sinh
R=\left(\frac{r_{-}^{2}-r^{2}}{r_{+}^{2}-r_{-}^{2}}\right)^{1/2}
\label{coshrin}
\eeq
we can construct the (3+2) global flat embedding for $r\leq r_{-}$
\bea
X^{0^{\prime}}&=&l\left(\frac{r_{-}^{2}-r^{2}}{r_{+}^{2}-r_{-}^{2}}\right)^{1/2}
       \sinh\left(\frac{r_{-}}{l^{2}}t-\frac{r_{+}}{l}\phi\right),\nonumber\\
X^{1^{\prime}}&=&l\left(\frac{r_{-}^{2}-r^{2}}{r_{+}^{2}-r_{-}^{2}}\right)^{1/2}
       \cosh\left(\frac{r_{-}}{l^{2}}t-\frac{r_{+}}{l}\phi\right), \nonumber \\
X^{0}&=&l\left(\frac{r_{+}^{2}-r^{2}}{r_{+}^{2}-r_{-}^{2}}\right)^{1/2}
        \cosh
        \left(\frac{r_{+}}{l^{2}}t-\frac{r_{-}}{l}\phi\right), \nonumber \\
X^{1}&=&l\left(\frac{r_{+}^{2}-r^{2}}{r_{+}^{2}-r_{-}^{2}}\right)^{1/2}
        \sinh
        \left(\frac{r_{+}}{l^{2}}t-\frac{r_{-}}{l}\phi\right), \nonumber \\
X^{2}&=&l.
\label{btz3}
\eea

Next, similar to the above cases, with the Killing vector
$\xi=\partial_{t}-N^{\phi}\partial_{\phi}$ we evaluate the three acceleration \beq
a_{3}=\frac{r_{+}^{2}r_{-}^{2}-r^{4}}{r^{2}l(r_{+}^{2}-r^{2})^{1/2}
(r_{-}^{2}-r^{2})^{1/2}}, \eeq and the Hawking temperature
\beq
T_{H}=\frac{a_{4}}{2\pi}=\frac{r(r_{+}^{2}-r_{-}^{2})}{2\pi
r_{+}l(r_{+}^{2}-r^{2})^{1/2} (r_{-}^{2}-r^{2})^{1/2}}.
\label{temp3}
\eeq

This completes the full global flat embeddings of the BTZ black hole in the two-time physics
scheme.  Note that the above (3+2) BTZ embedding solutions are
consistent with those in~\cite{btz2} where they obtained the (2+2) BTZ embeddings
in the standard global flat embedding scheme, without considering the Sp(2) local symmetry
(\ref{xmxmpm}) and the corresponding SO(3,2) global symmetry
invariant Lagrangian construction, which will be discussed  in the two-time physics
approach~\cite{bars98} in the next section.

\section{SO(3,2)/Sp(2) symmetries}
\setcounter{equation}{0}
\renewcommand{\theequation}{\arabic{section}.\arabic{equation}}

\subsection{Case I: $r\geq r_{+}$}

Now, in the global flat embedding (\ref{btz22}) for $r\ge r_{+}$, we
consider the Lorentz generators of the SO(3,2) symmetry \beq
L^{MN}=X^{M}P^{N}-X^{N}P^{M},\label{lmn} \eeq which, using the
conjugate pair $(X^{M},P^{M})$ in (\ref{xm1}) and
(\ref{pm1}), yields \bea L^{2M}&=& l P^{M},~~~
L^{0^{\prime}1^{\prime}}=p_{-},~~~
L^{01}=p_{+},\nn\\
L^{0^{\prime}0}&=&c_{-}s_{+}p_{R}-c_{-}c_{+}p_{+}{\rm
coth}R-s_{-}s_{+}p_{-}{\rm tanh}R,\nn\\
L^{1^{\prime}1}&=&s_{-}c_{+}p_{R}-s_{-}s_{+}p_{+}{\rm
coth}R-c_{-}c_{+}p_{-}{\rm tanh}R,\nn\\
L^{0^{\prime}1}&=&c_{-}c_{+}p_{R}-c_{-}s_{+}p_{+}{\rm
coth}R-s_{-}c_{+}p_{-}{\rm tanh}R,\nn\\
L^{1^{\prime}0}&=&s_{-}s_{+}p_{R}-s_{-}c_{+}p_{+}{\rm
coth}R-c_{-}s_{+}p_{-}{\rm tanh}R, \label{lmn11}
\eea
where we have used $\cosh R$ and $\sinh R$ in (\ref{coshout}).  Note
that these Lorentz generators produce the classical SO(3,2) global
symmetry transformations under $\delta$ defined as the Poisson
bracket $\delta=\frac{1}{2}\epsilon_{MN}\{L^{MN},~\}$
as follows\footnote{In the literature~\cite{btz2}, the Lorentz generators of the
SO(2,2) subgroup are constructed in the standard global flat embedding scheme.}
\bea
\delta\left(\frac{x^{+}}{l}\right)&=&(\epsilon_{0^{\prime}0}c_{-}c_{+}
+\epsilon_{0^{\prime}1}c_{-}s_{+}
+\epsilon_{1^{\prime}0}s_{-}c_{+}
+\epsilon_{1^{\prime}1}s_{-}s_{+}){\rm coth} R\nn\\
& &-\epsilon_{01} -(\epsilon_{02}c_{+}+\epsilon_{12}s_{+}){\rm
cosech} R, \nn\\
\delta R&=&-\epsilon_{0^{\prime}0}c_{-}s_{+}
-\epsilon_{0^{\prime}1}c_{-}c_{+}
-\epsilon_{1^{\prime}0}s_{-}s_{+}
-\epsilon_{1^{\prime}1}s_{-}c_{+}\nn\\
& &+(\epsilon_{0^{\prime}2}c_{-}
+\epsilon_{1^{\prime}2}s_{-})\sinh R +(\epsilon_{02}s_{+}
+\epsilon_{12}c_{+})\cosh R,\nn\\
\delta
\left(\frac{x^{-}}{l}\right)&=&(\epsilon_{0^{\prime}0}s_{-}s_{+}
+\epsilon_{0^{\prime}1}s_{-}c_{+}
+\epsilon_{1^{\prime}0}c_{-}s_{+}
+\epsilon_{1^{\prime}1}c_{-}c_{+}){\rm tanh} R\nn\\
& &-\epsilon_{0^{\prime}1^{\prime}}
+(\epsilon_{0^{\prime}2}s_{-}+\epsilon_{1^{\prime}2}c_{-}){\rm
sech} R. \label{delta11} \eea

On the other hand, in the two-time physics the SO(3,2) Lagrangian is
given by
\beq L=\dot{X}^{M}P_{M}-\frac{1}{2}A_{22}P^{M}P_{M},
\label{twotimelag}
\eeq
which, exploiting the conjugate pair $(X^{M},P^{M})$ in (\ref{xm1})
and (\ref{pm1}), yields
\beq
L=\frac{\dot{x}^{+}}{l}p_{+}+\dot{R}p_{R}+\frac{\dot{x}^{-}}{l}p_{-}
-\frac{1}{2}A_{22}~g^{\mu\nu}p_{\mu}p_{\nu}.
\label{laggmunu}
\eeq
Here the conjugate momenta $p_{\mu}$ is defined in (\ref{xpmunu}) and the metric
$g^{\mu\nu}$ is given by
\beq
g^{\mu\nu}=\frac{1}{l^{2}}{\rm diag}~(-{\rm cosech}^{2}R,1,{\rm
sech}^{2}R).
\label{gmunu11}
\eeq
The above Lagrangian (\ref{laggmunu}) can also be rewritten as
\beq
L=\frac{1}{2A_{22}}g_{\mu\nu}\dot{x}^{\mu}\dot{x}^{\nu},
\label{laggmunu2} \eeq with $x^{\mu}$ defined in (\ref{xpmunu}) and the inverse
metric $g_{\mu\nu}$.

After some algebra, we obtain the transformation rule for the
$A_{22}$ as below \beq \delta
A_{22}=2A_{22}[(\epsilon_{0^{\prime}2}c_{-}
+\epsilon_{1^{\prime}2}s_{-})\cosh R +(\epsilon_{02}s_{+}
+\epsilon_{12}c_{+})\sinh R],\label{dela22}\eeq from which,
together with the above transformation rules (\ref{delta11}), one
can easily see that the Lagrangian (\ref{twotimelag}) is SO(3,2) global symmetry
invariant.

\subsection{Case II: $r_{-}\le r\le r_{+}$}

Next, we consider the global flat embedding (\ref{btz222}) for $r_{-}\le
r\le r_{+}$.  As in the previous case, constructing $P^{M}$ conjugate to $X^{M}$
in (\ref{xm2}) as below
\bea
P^{M}&=&\frac{1}{l}(c_{-}p_{R}\sin R +s_{-}p_{-}{\rm sec}
R,s_{-}p_{R}\sin R+c_{-}p_{-}{\rm sec}R,
\nonumber\\
& &-c_{+}p_{R}\cos R+s_{+}p_{+}{\rm cosec} R,-s_{+}p_{R}\cos R
+c_{+}p_{+}{\rm cosec}R,0),\nn\\
\label{pm2} \eea
which satisfies the Sp(2) local symmetry (\ref{xmxmpm}), we can obtain the Lorentz
generators of the SO(3,2) global symmetry
\bea
L^{2M}&=& l P^{M},~~~
L^{0^{\prime}1^{\prime}}=p_{-},~~~
L^{01}=p_{+},\nn\\
L^{0^{\prime}0}&=&-c_{-}c_{+}p_{R}+c_{-}s_{+} p_{+} {\rm
cot}R-s_{-}c_{+}p_{-}{\rm tan}R,\nn\\
L^{1^{\prime}1}&=&-s_{-}s_{+}p_{R}+s_{-}c_{+}p_{+} {\rm
cot}R-c_{-}s_{+}p_{-}{\rm tan}R,\nn\\
L^{0^{\prime}1}&=&-c_{-}s_{+}p_{R}+c_{-}c_{+}p_{+} {\rm
cot}R-s_{-}s_{+}p_{-}{\rm tan}R,\nn\\
L^{1^{\prime}0}&=&-s_{-}c_{+}p_{R}+s_{-}s_{+}p_{+} {\rm
cot}R-c_{-}c_{+}p_{-}{\rm tan}R,
\label{lmn22}
\eea
to yield the classical SO(3,2) global symmetry transformations
for $x^{\mu}$ in (\ref{xpmunu})
\bea
\delta\left(\frac{x^{+}}{l}\right)&=&-(\epsilon_{0^{\prime}0}c_{-}s_{+}
+\epsilon_{0^{\prime}1}c_{-}c_{+}
+\epsilon_{1^{\prime}0}s_{-}s_{+}
+\epsilon_{1^{\prime}1}s_{-}c_{+}){\rm cot} R\nn\\
& &-\epsilon_{01} +(\epsilon_{02}s_{+}+\epsilon_{12}c_{+}){\rm
cosec} R, \nn\\
\delta R&=&\epsilon_{0^{\prime}0}c_{-}c_{+}
+\epsilon_{0^{\prime}1}c_{-}s_{+}
+\epsilon_{1^{\prime}0}s_{-}c_{+}
+\epsilon_{1^{\prime}1}s_{-}s_{+}\nn\\
& &+(\epsilon_{0^{\prime}2}c_{-} +\epsilon_{1^{\prime}2}s_{-})\sin
R -(\epsilon_{02}c_{+}
+\epsilon_{12}s_{+})\cos R,\nn\\
\delta
\left(\frac{x^{-}}{l}\right)&=&(\epsilon_{0^{\prime}0}s_{-}c_{+}
+\epsilon_{0^{\prime}1}s_{-}s_{+}
+\epsilon_{1^{\prime}0}c_{-}c_{+}
+\epsilon_{1^{\prime}1}c_{-}s_{+}){\rm tan} R\nn\\
& &-\epsilon_{0^{\prime}1^{\prime}}
+(\epsilon_{0^{\prime}2}s_{-}+\epsilon_{1^{\prime}2}c_{-}){\rm
sec} R. \label{delta22} \eea

On the other hand, inserting the conjugate pair $(X^{M},P^{M})$ in (\ref{xm2}) and
(\ref{pm2}) into the Lagrangian (\ref{twotimelag}), we obtain the metric
\beq
g^{\mu\nu}=\frac{1}{l^{2}}{\rm diag}~({\rm cosec}^{2}R,-1,{\rm
sec}^{2}R),\label{gmunu22} \eeq
from which we also obtain the SO(3,2) global symmetry transformations for $A_{22}$
\beq
\delta A_{22}=2A_{22}[(\epsilon_{0^{\prime}2}c_{-}
+\epsilon_{1^{\prime}2}s_{-})\cos R +(\epsilon_{02}c_{+}
+\epsilon_{12}s_{+})\sin R].
\label{dela222}
\eeq
Exploiting the transformation rules (\ref{delta22}) and (\ref{dela222}), one
can easily see that the Lagrangian (\ref{twotimelag}) with $X^{M}$ and $P^{M}$
in (\ref{xm2}) and (\ref{pm2}) is invariant under the SO(3,2) global
symmetry transformations.

\subsection{Case III: $r\leq r_{-}$}

Finally, we consider the (3+2) global flat embedding for $r\leq r_{-}$.
Exploiting $X^{M}$ in (\ref{btz3}) and constructing its conjugate momenta $P^{M}$
as below
\bea
P^{M}&=&\frac{1}{l}(s_{-}p_{R}\cosh R -c_{-} p_{-}{\rm cosech}R,
c_{-}p_{R}\cosh R-s_{-}p_{-}{\rm cosech}R,
\nonumber\\
& &c_{+}p_{R}\sinh R+s_{+}p_{+}{\rm sech}R,s_{+}p_{R}\sinh R
+c_{+}p_{+}{\rm sech}R,0),\label{pm3}
\eea
to satisfy the Sp(2) symmetry (\ref{xmxmpm}), we can obtain the Lorentz
generators of the SO(3,2) symmetry \bea L^{2M}&=& l P^{M},~~~
L^{0^{\prime}1^{\prime}}=p_{-},~~~
L^{01}=p_{+},\nn\\
L^{0^{\prime}0}&=&-s_{-}c_{+}p_{R}+s_{-}s_{+}
 p_{+}{\rm tanh}R
+c_{-}c_{+}p_{-}{\rm coth}R,\nn\\
L^{1^{\prime}1}&=&-c_{-}s_{+}p_{R}+c_{-}c_{+}p_{+}{\rm
tanh}R+s_{-}s_{+}p_{-}{\rm coth}R,\nn\\
L^{0^{\prime}1}&=&-s_{-}s_{+}p_{R}+s_{-}c_{+}p_{+}{\rm
tanh}R+c_{-}s_{+}p_{-}{\rm coth}R,\nn\\
L^{1^{\prime}0}&=&-c_{-}c_{+}p_{R}+c_{-}s_{+}p_{+}{\rm
tanh}R+s_{-}c_{+} p_{-}{\rm coth}R.
\label{lmn33}
\eea
Inserting the Lorentz generators (\ref{lmn33}) into the classical SO(3,2) global
symmetry transformation rules $\delta=\frac{1}{2}\epsilon_{MN}\{L^{MN},~\}$
yields
\bea \delta
\left(\frac{x^{+}}{l}\right)&=&-(\epsilon_{0^{\prime}0}s_{-}s_{+}
+\epsilon_{0^{\prime}1}s_{-}c_{+}
+\epsilon_{1^{\prime}0}c_{-}s_{+}
+\epsilon_{1^{\prime}1}c_{-}c_{+}){\rm tanh} R\nn\\
& &-\epsilon_{01} +(\epsilon_{02}s_{+}+\epsilon_{12}c_{+}){\rm
sech} R, \nn\\
\delta R&=&\epsilon_{0^{\prime}0}s_{-}c_{+}
+\epsilon_{0^{\prime}1}s_{-}s_{+}
+\epsilon_{1^{\prime}0}c_{-}c_{+}
+\epsilon_{1^{\prime}1}c_{-}s_{+}\nn\\
& &+(\epsilon_{0^{\prime}2}s_{-}
+\epsilon_{1^{\prime}2}c_{-})\cosh R +(\epsilon_{02}c_{+}
+\epsilon_{12}s_{+})\sinh R,\nn\\
\delta
\left(\frac{x^{-}}{l}\right)&=&-(\epsilon_{0^{\prime}0}c_{-}c_{+}
+\epsilon_{0^{\prime}1}c_{-}s_{+}
+\epsilon_{1^{\prime}0}s_{-}c_{+}
+\epsilon_{1^{\prime}1}s_{-}s_{+}){\rm coth} R\nn\\
& &-\epsilon_{0^{\prime}1^{\prime}}
-(\epsilon_{0^{\prime}2}c_{-}+\epsilon_{1^{\prime}2}s_{-}){\rm
cosech} R. \label{delta33}
\eea
On the other hand, substituting the conjugate pair
$(X^{M},P^{M})$ in (\ref{xm3}) and (\ref{pm3}) into the Lagrangian
(\ref{twotimelag}), we obtain the metric
\beq
g^{\mu\nu}=\frac{1}{l^{2}}{\rm diag}~({\rm sech}^{2}R,1,-{\rm
cosech}^{2}R),\label{gmunu33}
\eeq
so that we can construct the SO(3,2) global symmetry transformations for $A_{22}$
\beq \delta
A_{22}=2A_{22}[(\epsilon_{0^{\prime}2}s_{-}
+\epsilon_{1^{\prime}2}c_{-})\sinh R +(\epsilon_{02}c_{+}
+\epsilon_{12}s_{+})\cosh R].
\label{dela2222}
\eeq
In the region inside the inner horizon where $X^{M}$ and $P^{M}$ are given by (\ref{btz3})
and (\ref{pm3}), as in the previous cases, we can thus obtain the SO(3,2) global symmetry
invariant Lagrangian under the transformation rules (\ref{delta33}) and (\ref{dela2222}).

Now it seems appropriate to comment on the Sp(2) local symmetry associated with the two-time physics,
for the BTZ global embedding solutions.  To be more specific, we consider $X^{M}$ and $P^{M}$ for
$r\ge r_{+}$ in (\ref{xm1}) and (\ref{pm1})  to evaluate
\beq
P_{M}P^{M}=l^{-2}(-p_{+}^{2}{\rm cosech}^{2}R+p_{R}^{2}+p_{-}^{2}{\rm sech}^{2}R).
\eeq
Exploiting the metric $g^{\mu\nu}$ in (\ref{gmunu11}), we can rewrite $p_{\mu}$ in (\ref{xpmunu})
in terms of the $\dot{x}^{\mu}$
\beq
(p^{+},p_{R},p^{-})=\frac{l^{2}}{A_{22}}\left(-\frac{\dot{x}^{+}}{l}\sinh^{2}R,\dot{R},
\frac{\dot{x}^{-}}{l}\cosh^{2}R\right),
\eeq
to, together with the explicit expressions for $\cosh R$ and $\sinh R$ in (\ref{coshout}),
yield a relation between the $P_{M}P^{M}$ and the BTZ metric
\beq
P_{M}P^{M}=\frac{1}{A_{22}^{2}}\left(\frac{ds}{dt}\right)^{2},
\label{ppdsdt}
\eeq
which vanishes since the BTZ metric or line element itself is time-independent.  We can thus
explicitly show that the Sp(2) symmetry (\ref{xmxmpm}) is conserved in the (3+2) BTZ global
embedding solutions.  Moreover, the Lagrangian (\ref{twotimelag}) can be rewritten as
\beq
L=\frac{1}{2}A_{22}P_{M}P^{M}.
\eeq
Here one notes that, due to the BTZ metric time-independence, the SO(3,2) Lagrangian or Hamiltonian
vanishes, which is a characteristic of the two-time physics~\cite{bars98}.

\section{Conclusions}
\setcounter{equation}{0}
\renewcommand{\theequation}{\arabic{section}.\arabic{equation}}

In conclusion, in the framework of two-time physics scheme, we have explicitly obtained
the global flat embeddings, three accelerations and Hawking temperatures of the BTZ black
holes both inside and outside the event horizons by exploiting the Sp(2) local symmetry.
Moreover, we have constructed
the SO(3,2) global symmetry invariant Lagrangians associated with these  BTZ black hole
embedding solutions.  The Sp(2) local symmetry has been also discussed in terms of the
metric time-independence.

\acknowledgments The author would like to thank Itzhak Bars for initial discussions. This
work is supported in part by the Korea Science and Engineering Foundation Grant
R01-2000-00015.

\appendix
\section{}
\setcounter{equation}{0}
\renewcommand{\theequation}{A.\arabic{equation}}

Now we consider the charged BTZ black hole with 3-metric \beq
ds^{2}=-N^{2}dt^{2}+N^{-2}dr^{2}+r^{2}d\phi^{2}, \label{3metric}
\eeq where the charged lapse function is given
as~\cite{btz1,btz2,cal} \beq
N^{2}=-M+\frac{r^{2}}{l^{2}}-2q^{2}\ln r, \label{lapsebtz} \eeq
where the mass $M$ can be rewritten as
$M=r_{H}^{2}/l^{2}-2q^{2}\ln r_{H}$ with the horizon $r_{H}(q)$,
which is the root of  $-M+r^{2}/l^{2} -2q^{2}\ln r=0$.  The
surface gravity in this charged BTZ black hole is given by
$k_{H}=[(r_{H}/l)^{2}-q^{2}]/r_{H}$. Making an ansatz of four
coordinates $(z^{0}, z^{1},z^{2},z^{3})$ in (\ref{zzzbtzc}),
which is different from the previous one in~\cite{kps00}, we
can construct the (3+2) global flat embedding
$ds^{2}=-(dz^{0})^2+(dz^{1})^2+(dz^{2})^2+(dz^{3})^2-(dz^{4})^2$
with the following coordinate transformations for $ql\le r_{H}<r$
\bea z^{0}&=&k_{H}^{-1}\left(-M+\frac{r^{2}}{l^{2}}-2q^{2}\ln
r\right)^{1/2}
\sinh k_{H}t, \nonumber \\
z^{1}&=&k_{H}^{-1}\left(-M+\frac{r^{2}}{l^{2}}-2q^{2}\ln r\right)^{1/2}
\cosh k_{H}t, \nonumber \\
z^{2}&=&r\cos\phi, \nonumber \\
z^{3}&=&r\sin\phi, \nonumber \\
z^{4}&=&\int dr~\left[
\frac{(1+q^{2}l^{2}/rr_{H})(1-q^{2}l^{2}/rr_{H})}{k_{H}^{2}l^{2}
[1-(q^{2}l^{2}/r_{H}^{2})n(r,r_{H})]}+1\right]^{1/2}.
\label{zzzbtzc}
\eea
Here $n(r,r_{H})$ is given by
\beq
n(r,r_{H})=\frac{2r_{H}^{2}}{r^{2}-r_{H}^{2}}\ln \frac{r}{r_{H}},
\eeq
which, due to L'Hostpital's rule, approaches unity as $r$ goes to $r_{H}$.

\end{document}